\documentclass[preprint,showpacs,12pt]{revtex4}

\usepackage{amsmath,amssymb,amsfonts}
\usepackage{graphicx}
\usepackage{hhline}
\usepackage{bm}
\usepackage{amsmath}
\usepackage{epsfig}
\usepackage{graphicx}

\usepackage{tabularx}
\usepackage{lscape}
\usepackage{rotating}
\usepackage{pstricks}

\usepackage{footnote}

\usepackage{ulem}

\begin{document}
\title{Investigation of the MicroBooNE neutrino cross sections on Argon}

\author{M. Martini}
\affiliation{IPSA-DRII,  63 boulevard de Brandebourg, 94200 Ivry-sur-Seine, France}
\affiliation{Sorbonne Universit\'e, Universit\'e Paris Diderot, CNRS/IN2P3, Laboratoire
de Physique Nucl\'eaire et de Hautes Energies (LPNHE), Paris, France}
\author{M. Ericson} 
\affiliation{Univ Lyon, Univ Claude Bernard Lyon 1, CNRS/IN2P3, IP2I Lyon, UMR 5822, F-69622, Villeurbanne, France}
\affiliation{Theory division, CERN, CH-12111 Geneva }
\author{G. Chanfray} 
\affiliation{Univ Lyon, Univ Claude Bernard Lyon 1, CNRS/IN2P3, IP2I Lyon, UMR 5822, F-69622, Villeurbanne, France}

\begin{abstract}
Experimental data of charged current inclusive neutrino cross sections on argon as a function of different variables have recently appeared.  
We have compared them to our theoretical approach. 
Overall we find an agreement in spite of a tendency of underestimation in some specific regions.  
A new aspect is the availability of data in terms of the energy transfer to the nucleus, which allows a better separation of the different reaction mechanisms.  
We explain the deterioration of agreement in specific kinematical conditions by the absence in our model of two-pion production and other inelastic channels, more important for MicroBooNE than for T2K. 
\end{abstract}

\maketitle

\section{Introduction}
In order to increase the number of interactions, neutrino experiments at intermediate energies use
complex nuclei as detectors. 
Following the MiniBooNE measurement \cite{AguilarArevalo:2010zc} of charged current quasielastic-like (CCQE-like) cross section, 
a large number of data has been collected, mainly on carbon (the target nucleus of MiniBooNE, MINERvA, T2K and NOvA experiments), 
but also on oxygen and iron. For a review see for example Refs. \cite{Katori:2016yel, Alvarez-Ruso:2017oui}. 

Another important nucleus is argon. Argon detectors are used in the context of the short-baseline Fermilab neutrino program \cite{Machado:2019oxb} aimed at the investigation of the possible 
 existence of a sterile neutrino, a topic recently revived by new MicroBooNE measurements \cite{MicroBooNE:2021rmx,MicroBooNE:2021sne,MicroBooNE:2021jwr}. Argon nucleus will also be used in DUNE, 
one of the next-generation long-baseline neutrino oscillation experiments \cite{Branca:2021vis}. Recently, after the pioneering measurements by ArgoNeuT \cite{ArgoNeuT:2011bms}, four  
muon neutrino differential 
cross sections on argon have been published by the MicroBooNE collaboration: 
two inclusive measurements \cite{Abratenko:2019jqo,MicroBooNE:2021cue} that include all charged-current events in which one muon is detected and two semi-inclusive ones, characterized by the detection of zero pion, one muon and one proton \cite{Abratenko:2020acr} and by the detection of zero pion, one muon and any number of protons \cite{Abratenko:2020sga}, respectively called CC0$\pi$1p and CC0$\pi$Np.
These results refer to flux-integrated cross sections. 
The flux involved is the Booster Neutrino Beam (BNB), published in Ref.~\cite{Abratenko:2019jqo} 
for the MicroBooNE experiment. Its normalized behavior  is very similar to the previous MiniBooNE one~\cite{AguilarArevalo:2008yp}. 

The aim of the present paper is to confront the  predictions of our approach on neutrino interactions with nuclei with the MicroBooNE experimental 
data on argon, focusing on the inclusive ones. 
Other theoretical evaluations of neutrino cross sections on argon have been performed \cite{Meucci:2013gja,Gallmeister:2016dnq,VanDessel:2017ery,Akbar:2017dih,VanDessel:2019atx,Nikolakopoulos:2020alk,Barbaro:2021psv,Franco-Patino:2021yhd,Butkevich:2021sfn,MicroBooNE:2021ccs,Gonzalez-Rosa:2022ltp}.

\section{Our theoretical model}
Our theoretical model has been described in detail in Ref.~\cite{Martini:2009uj}. It is based on nuclear response functions treated
in the random phase approximation (RPA) on top of a local Fermi gas calculation. It includes the
quasielastic cross section, the multinucleon emission and the coherent
and incoherent single pion production. 
In our works~\cite{Martini:2009uj,Martini:2010ex} we explained the debated MiniBooNE quasielastic-like cross section on carbon~\cite{AguilarArevalo:2010zc} by stressing the crucial role of the multinucleon (also called 2p-2h or np-nh) excitations contribution which sizeably increases the genuine quasielastic cross section.
We have made in Ref.~\cite{Martini:2011wp} an improvement of our original description of the quasielastic channel with the introduction 
of relativistic corrections. Our model has successfully reproduced several muon-neutrino flux-integrated double differential cross sections: 
the MiniBooNE CCQE-like neutrino~\cite{Martini:2011wp} and antineutrino~\cite{Martini:2013sha} ones, 
as well as their combination~\cite{Ericson:2015cva}, the MiniBooNE CC1$\pi^+$ and the T2K CC inclusive~\cite{Martini:2014dqa}, 
the T2K CC0$\pi$ for neutrinos~\cite{Abe:2016tmq} as well as for  antineutrinos and their combination~\cite{Abe:2020jbf}.  
Single differential electron-neutrino T2K CC inclusive cross sections have been successfully reproduced as well \cite{Martini:2016eec}. 
The agreement of our predictions with several neutrino flux-integrated differential cross section data was, in some kinematical conditions (forward lepton scattering angle, low $Q^2$), the result of a delicate balance between the quenching of the quasielastic due to RPA effects and the multinucleon enhancement. In this context, it is important to remind that the amount of RPA effects is model dependent. When the starting point of the calculation is the local Fermi gas, as in our case, these effects can lead to large reduction of the quasielastic cross sections. On the contrary, when a proper nucleus ground state is considered, \textit{i.e.} when mean field potential and/or realistic spectral function are used, RPA effects are smaller \cite{Pandey:2014tza,Nieves:2017lij}. Despite the differences on the treatment of the nucleus ground state, and consequently on the amount of RPA effects, the results obtained for the quasielastic cross section are globally in agreement each others when RPA and/or spectral function effects are taken into account. This is shown in Ref. \cite{Nieves:2017lij} where the interplay between these two nuclear effects is analyzed in large details. The correspondence between the response function and the spectral function approaches is discussed also by us in Ref. \cite{Chanfray:2021tsp}, where we have shown that the amount of nucleon correlations of our approach is compatible with the one obtained via \textit{ab-initio} calculations.

All the studies mentioned above concerned the $^{12}$C cross sections. 
In the present article we want to  confront our theory to recent inclusive MicroBooNE data \cite{Abratenko:2019jqo,MicroBooNE:2021cue} which refer to $^{40}$Ar.  
This nucleus  has a different number of protons (18) and neutrons (22). 
In order to keep the description close to our previous one on $^{12}$C, 
we perform the calculations of nuclear responses, based on local density approximation, by approximating both the proton and the neutron density profiles of $^{40}$Ar by the proton density profile of $^{40}$Ca$^1$\footnotetext[1]{The impact of $^{40}$Ar Fermi momentum asymmetry for protons and neutrons on the quasielastic responses per nucleon has been investigated  by Barbaro \textit{et al.}~\cite{Barbaro:2018kxa} who find small effects. It may justify our approximation to calculate the responses per nucleon for the symmetric $^{40}$Ca.}. 
%
For the quasielastic cross section of $^{40}$Ar, we rescale the $^{40}$Ca results  according to the real number of active nucleons, \textit{i.e} the quasielastic  $^{40}$Ca results are multiplied by $\frac{22}{20}$. For one pion production we do not consider a possible rescaling effect which is less important since both protons and neutrons are active in this channel.  Concerning the 2p-2h excitations, for the quasi-deuteronic contribution, which is the dominant term and goes as the product of the proton and neutron densities, $\rho_p\rho_n$, the difference between argon and calcium represents only a $1\%$ correction. This is due to the fact that the excess of neutrons in argon with respect to calcium is equal to the deficit of protons. For the np-nh sector, in our approach we include the nucleon-nucleon (NN) short-range correlations, the 
$\Delta$-MEC contribution and the NN correlations-MEC interference, that we denote N$\Delta$ interference. In this work we calculate the $\Delta$-MEC contribution directly for $^{40}$Ca. Instead for the NN and N$\Delta$ interference parts, assuming a linear $A$-dependence, we simply rescale our previous parametrization for $^{12}$C 
(the so called ``new'' in Ref.\cite{Martini:2009uj}) 
by a factor $\frac{40}{12}$. For the NN contributions this assumption is reasonable, as discussed in Ref.~\cite{Mosel:2016uge}, even if it deviates by about 20\% from the full results obtained in Ref.~\cite{Mosel:2016uge} which take into account nuclear surface corrections: a ratio of about 4.2 for the presence of short-range pairs in $^{40}$Ar vs. $^{12}$C is predicted in Ref.~\cite{Mosel:2016uge}, while $40/12=3.\bar{3}$. 

Beyond the inclusive measurements \cite{Abratenko:2019jqo,MicroBooNE:2021cue}, the MicroBooNE collaboration also published two semi-inclusive $\nu_\mu$ cross sections: one with one muon, no pion and one single proton in the final state \cite{Abratenko:2020acr}, commonly referred as CC0$\pi$1p cross section, and another with 
one muon, zero pion and at least
one proton 
in the final state \cite{Abratenko:2020sga}, called CC0$\pi$Np cross section. 
Beyond the differences in the final states, 
the two measurements also differ on the restrictions applied to the muon and proton phase-space. 
In the case of CC0$\pi$1p, the proton kinematic constraints are 
$0.3<p_p<1.0$ GeV/c and $\cos\theta_p>0.15$. 
The proton kinematic constraints in the CC0$\pi$Np measurement are less stringent: $0.3<p_p<1.2$ GeV/c. 
Since our formalism is based on inclusive response functions, which are built by integrating on the nucleonic kinematical variables and depend only on the transferred energy $\omega$ and on the norm of the momentum transfer $|\vec{q}|$, we cannot constrain the kinematics of outgoing protons. Furthermore in our formalism we do not include final state interactions (FSI) for the proton neither pion absorption due to pion FSI. 
For these reasons our inclusive theoretical predictions cannot be directly compared with the semi-inclusive measurements and here we focus only on inclusive cross sections. The extension of our formalism to semi-inclusive processes requires further theoretical developments.

\section{Results on charged current inclusive cross sections }
\subsection{MicroBooNE and T2K double differential  cross sections}
\label{susec_ccincl_d2s} 
In the charged current interaction inclusive measurement only the charged lepton is detected and all reaction mechanisms 
 (quasielastic, nuclear resonances, multinucleon excitations, one- and multi-pion production, deep inelastic scattering) can contribute to this process. 
Since all channels are included, the inclusive measurements present two advantages with respect to exclusive channels measurements. 
Firstly they are not affected by the necessity to subtract the background which comes from the open channels other than the channel in consideration. Second they accumulate more rapidly enough statistics of events to allow more detailed cross sections evaluations. 
Indeed the first MicroBooNE published cross section is the inclusive one~\cite{Abratenko:2019jqo} 
and concerns double differential cross sections. 
While instead the subsequent exclusive results, CC0$\pi$1p~\cite{Abratenko:2020acr} and CC0$\pi$Np~\cite{Abratenko:2020sga}, 
refer to single differential cross sections. 

\begin{figure}
\begin{center}
  \includegraphics[width=8cm]{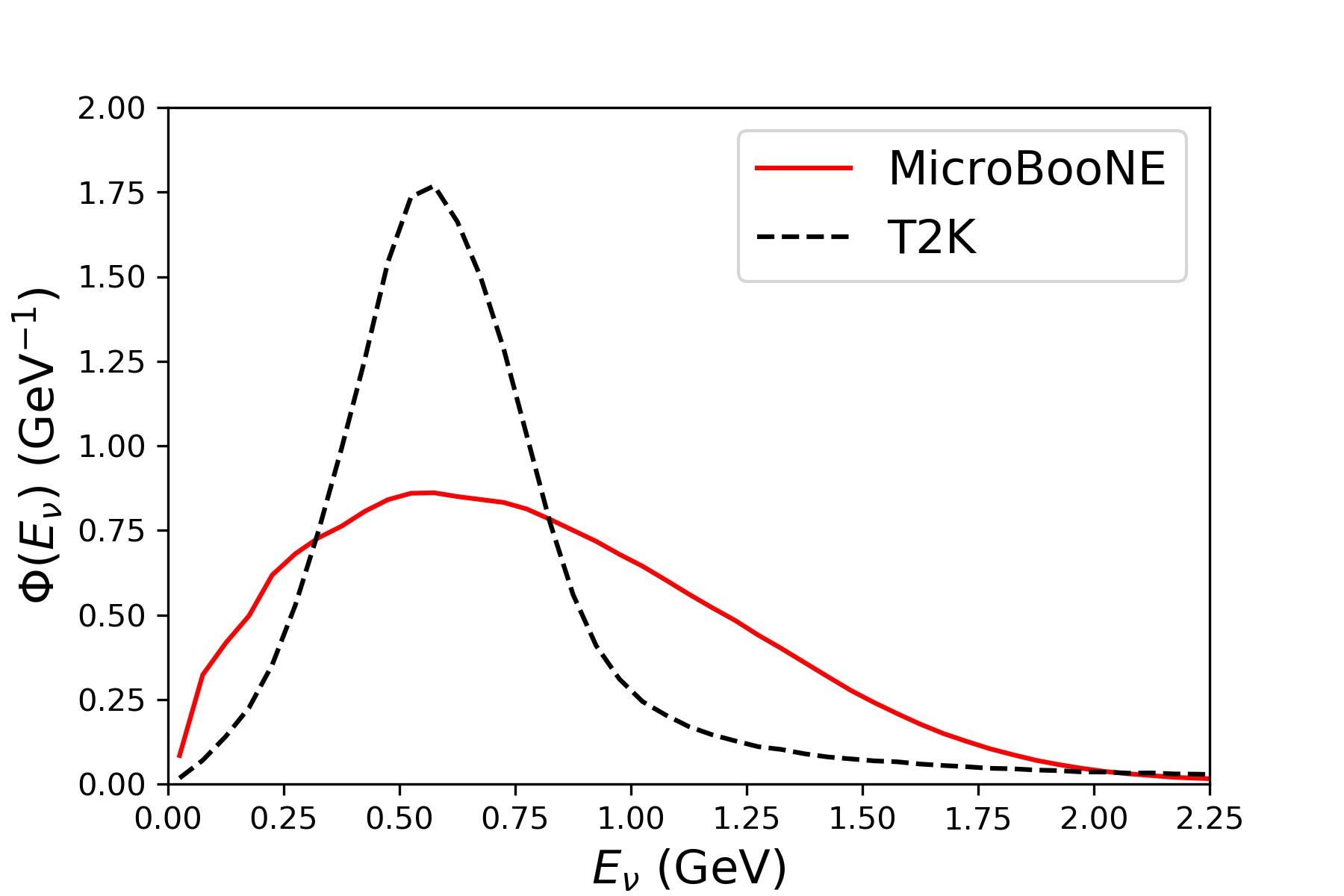} 
\caption{MicroBooNE and T2K normalized $\nu_\mu$ fluxes.}
\label{fig_micro_t2k_fluxes}
\end{center}
\end{figure}

Notice that in the flux-integrated inclusive cross section, 
the relative weight of the different mechanisms depend on the energy profile of the neutrino beam 
as different channels peak in different neutrino energy regions. 
In our present evaluation of the MicroBooNE flux-integrated double differential cross section we assume that the quasielastic, 
the multinucleon emission and the single pion production (coherent and incoherent) are the only open channels. 
With this assumption we have successfully reproduced ~\cite{Martini:2014dqa} the data of the T2K experiment 
whose neutrino flux peaks in the same region as the MicroBooNE one, but has a different energy profile, as shown in Fig. \ref{fig_micro_t2k_fluxes}. 
 
\begin{figure}
\begin{center}
  \includegraphics[width=16cm]{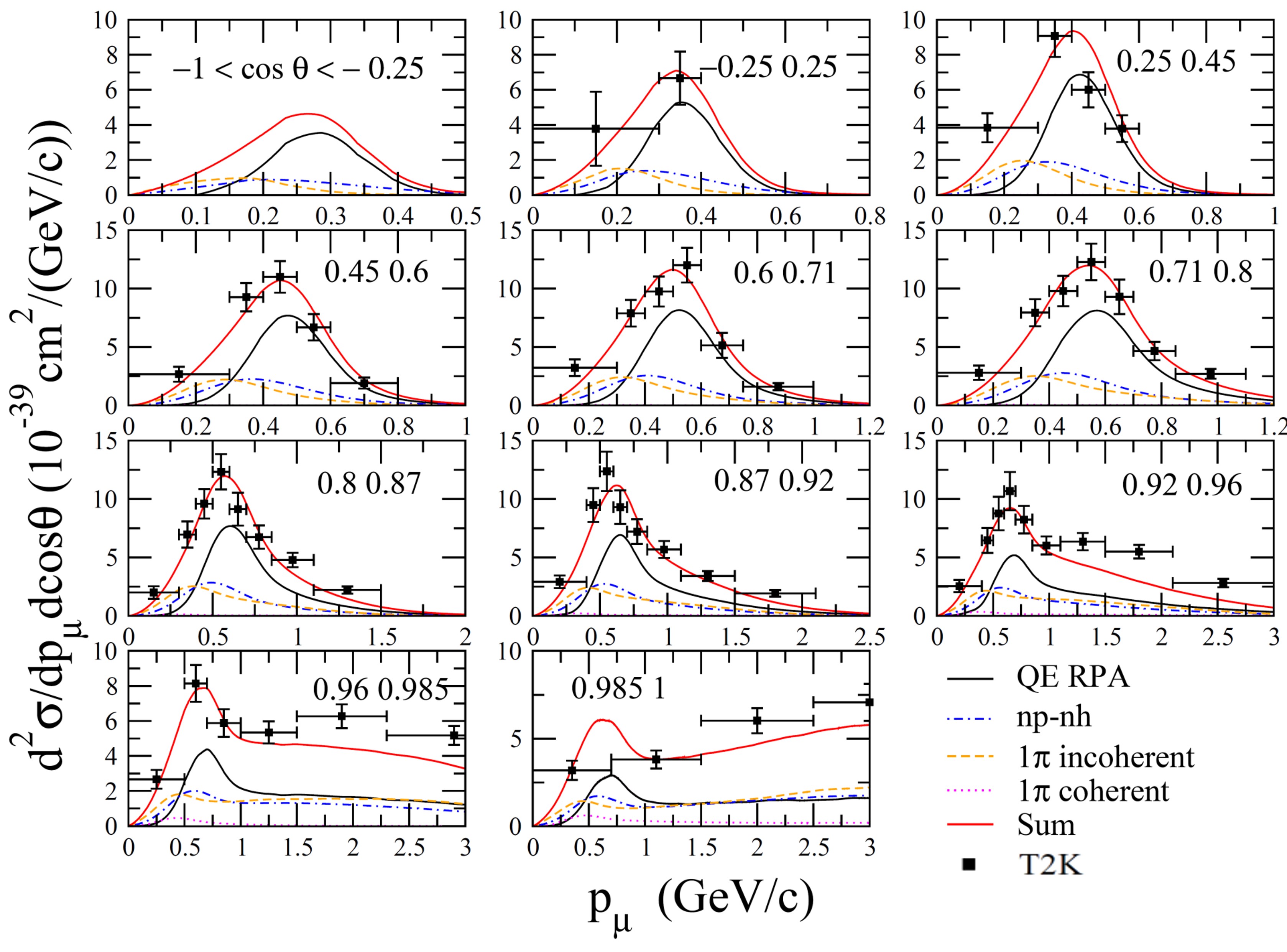}  
\caption{T2K flux-integrated $\nu_\mu$ CC inclusive 
double differential cross section per nucleon 
of carbon
as a function of the muon momentum $p_\mu$ calculated in average for different cosine intervals of muon scattering angle $\theta$.
The experimental T2K polystyrene data are taken from Ref.~\cite{Abe:2018uhf}. 
The different contributions to the inclusive cross sections obtained in our approach are shown.}
\label{fig_t2k_d2s_2018}
\end{center}
\end{figure}

Before we turn to the inclusive MicroBooNE cross section on argon, we investigate further the inclusive cross section per nucleon of carbon by considering more recent inclusive T2K data on polystyrene \cite{Abe:2018uhf}. They are characterized by an increased angular acceptance and higher statistics 
as compared to the previous ones of Ref.~\cite{Abe:2013jth} that we used as a test of our approach in Ref.~\cite{Martini:2014dqa}. 
The larger precision of these recent data allows a more stringent test of our model. In Fig. \ref{fig_t2k_d2s_2018} 
we compare our calculations with these data. Since the results are given per nucleon, the difference between our theoretical calculation on carbon and the experimental measurement, which involves carbon (86.1\%), hydrogen (7.4\%) and oxygen (3.7\%), can be ignored. The agreement is remarkable in nearly all the bins.  
Some deviation appears in the forward direction for $p_\mu>1.5$ GeV/c. A similar discrepancy appears in the comparison between the data and the Monte Carlo simulations performed by T2K. 

\begin{figure}
\begin{center}
  \includegraphics[width=16cm]{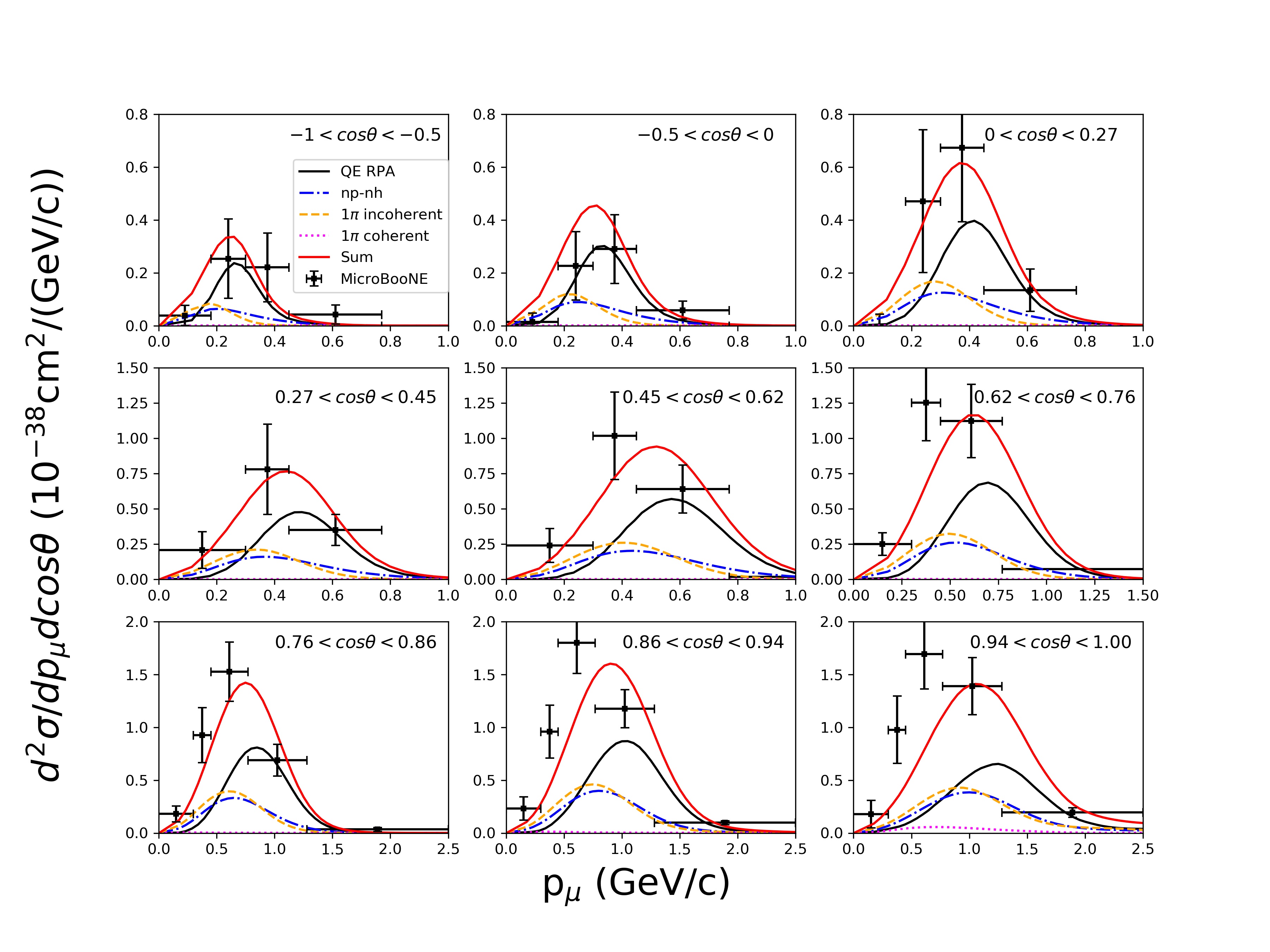}
\caption{MicroBooNE flux-integrated $\nu_\mu$ CC inclusive 
double differential cross section on argon per nucleon
as a function of the muon momentum $p_\mu$ calculated in average for different cosine intervals of muon scattering angle $\theta$.
The experimental MicroBooNE data are taken from Ref.~\cite{Abratenko:2019jqo}. 
The different contributions to the inclusive cross sections obtained in our model are shown.}
\label{fig_comp_inclusive_micro}
\end{center}
\end{figure}

We now turn to the MicroBooNE experiment. In Fig.\ref{fig_comp_inclusive_micro} we confront
our calculations with these data. 
We plot separately the different contributions to the inclusive cross section: the quasielastic, the multinucleon emission (np-nh) and the one pion production 
(coherent and incoherent).  
The overall agreement is reasonable, though  not as good as in the $^{12}$C T2K case: a disagreement shows up for low $p_\mu$. 
The SuSA calculation of  Ref.~\cite{Barbaro:2021psv} displays a similar trend. The same features appear in the recent SuSAv2 results~\cite{Gonzalez-Rosa:2022ltp} which incorporate inelastic contributions beyond Delta excitations, absent in our case.

\subsection{MicroBooNE energy dependent cross sections}
\label{susec_ccincl_en_dep}


At this stage it is interesting to analyze new MicroBooNE data on inclusive cross section \cite{MicroBooNE:2021cue} which may help to elucidate the reason of the disagreement at low $p_\mu$. The new measured quantities are the total cross section as a function of the neutrino energy $\sigma(E_\nu)$, 
the flux integrated differential cross section as a function of the muon energy $\frac{d\sigma}{d E_\mu}$, and the flux integrated differential cross section as a function of the transferred energy $\frac{d\sigma}{d \omega}$.  This is the first measurement of this type, after the MINERvA one \cite{MINERvA:2015ydy} at larger neutrino energies. 
We remind that, since the neutrino beams are not monochromatic, all the neutrino cross sections that are not expressed in terms of the measured lepton kinematics variables are affected by the energy reconstruction problem \cite{Martini:2012fa,Martini:2012uc,Nieves:2012yz,Lalakulich:2012hs,Ankowski:2016jdd}. This is the case of $\sigma(E_\nu)$ and $\frac{d\sigma}{d \omega}$. 
In Ref.\cite{MicroBooNE:2021cue} the experimental results have been presented for the first time as a function of true neutrino energy $E_\nu$ and transferred energy $\omega$. 
 This has been made possible by a new procedure (based on the comparison between the data and the Monte Carlo predictions constrained on the lepton kinematics) allowing the mapping between the true $E_\nu$ and $\omega$ on one hand, and the reconstructed neutrino energy $E_\nu^{\textrm{rec}}$ and hadronic energy $E_{\textrm{had}}^{\textrm{rec}}$ on the other hand. 
This new way to present the experimental results allows us to compare the data with our theoretical results as a function of the true $E_\nu$ and $\omega$.

\begin{figure}
\begin{center}
  \includegraphics[width=12cm]{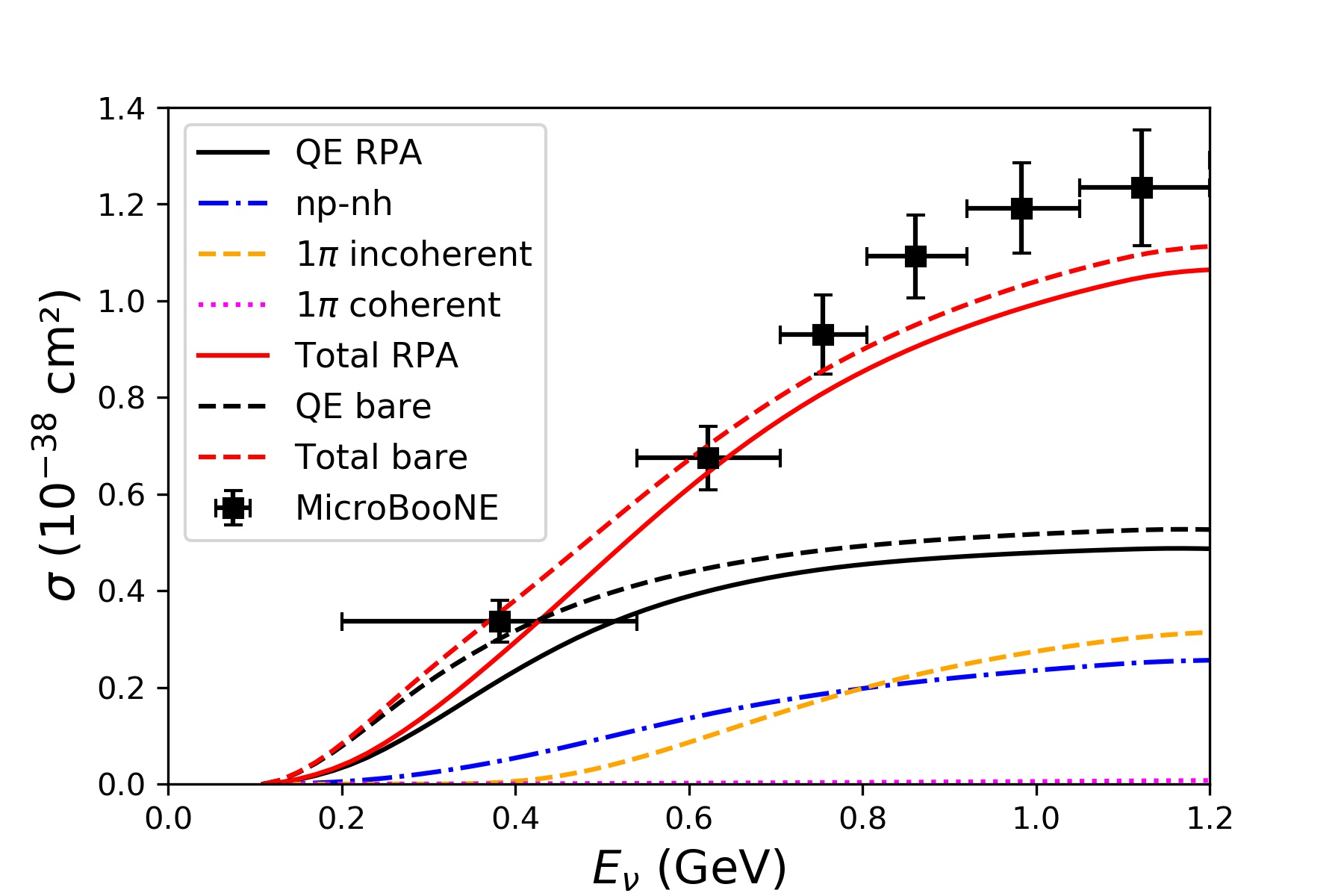}
\caption{Charged current inclusive $\nu_\mu$ cross
section on argon per nucleon as a function of the neutrino energy. The experimental MicroBooNE data are taken from Ref.~\cite{MicroBooNE:2021cue}.
The genuine quasielastic, the np-nh contributions, the coherent and incoherent 1$\pi$ production as well as the sum of these channels calculated in our RPA model are shown. The genuine quasielastic and the total cross section are shown also when the RPA is switched off.}
\label{fig_sigma_vs_Enu}
\end{center}
\end{figure}

\begin{figure}
\begin{center}
  \includegraphics[width=8cm]{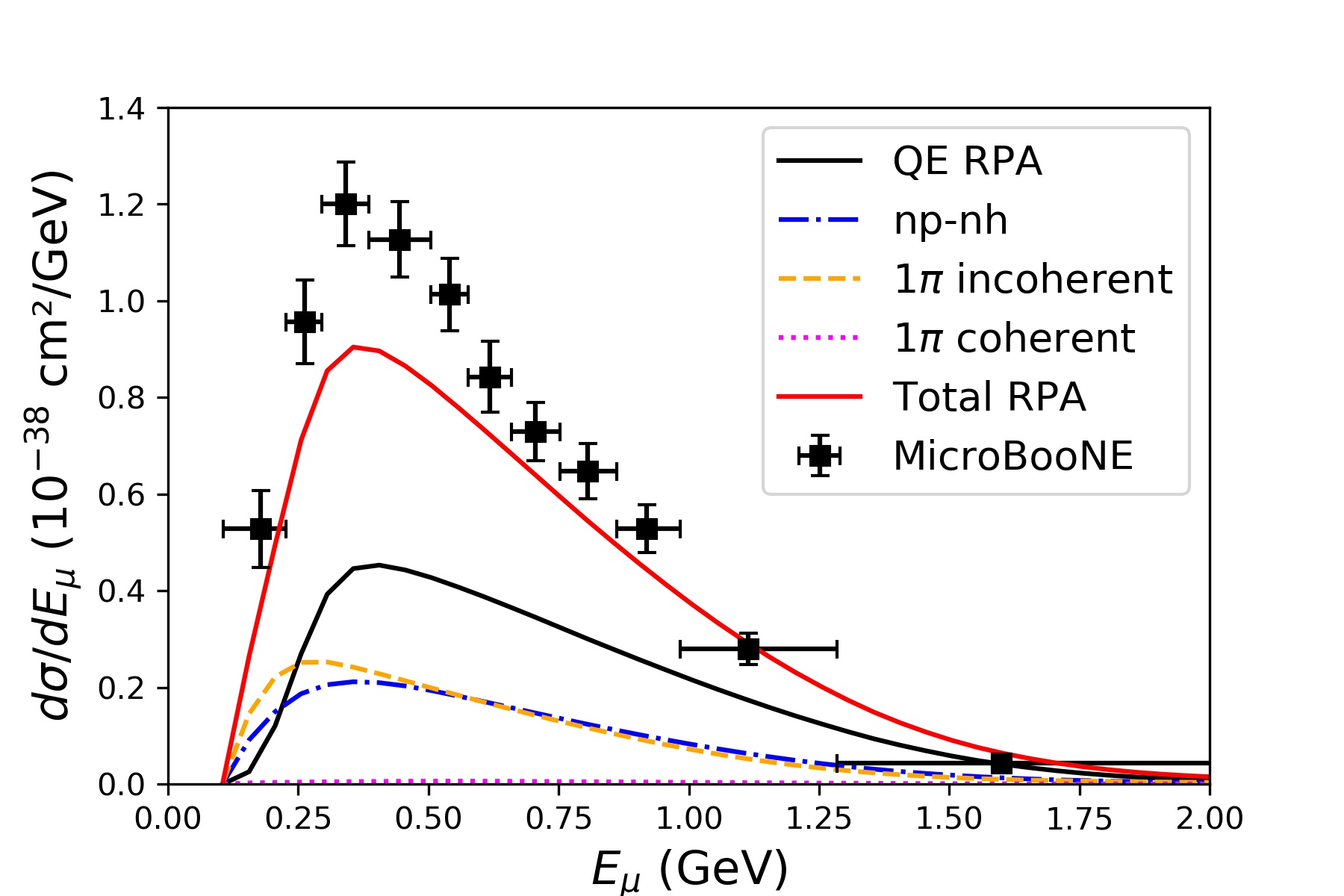}
  \includegraphics[width=8cm]{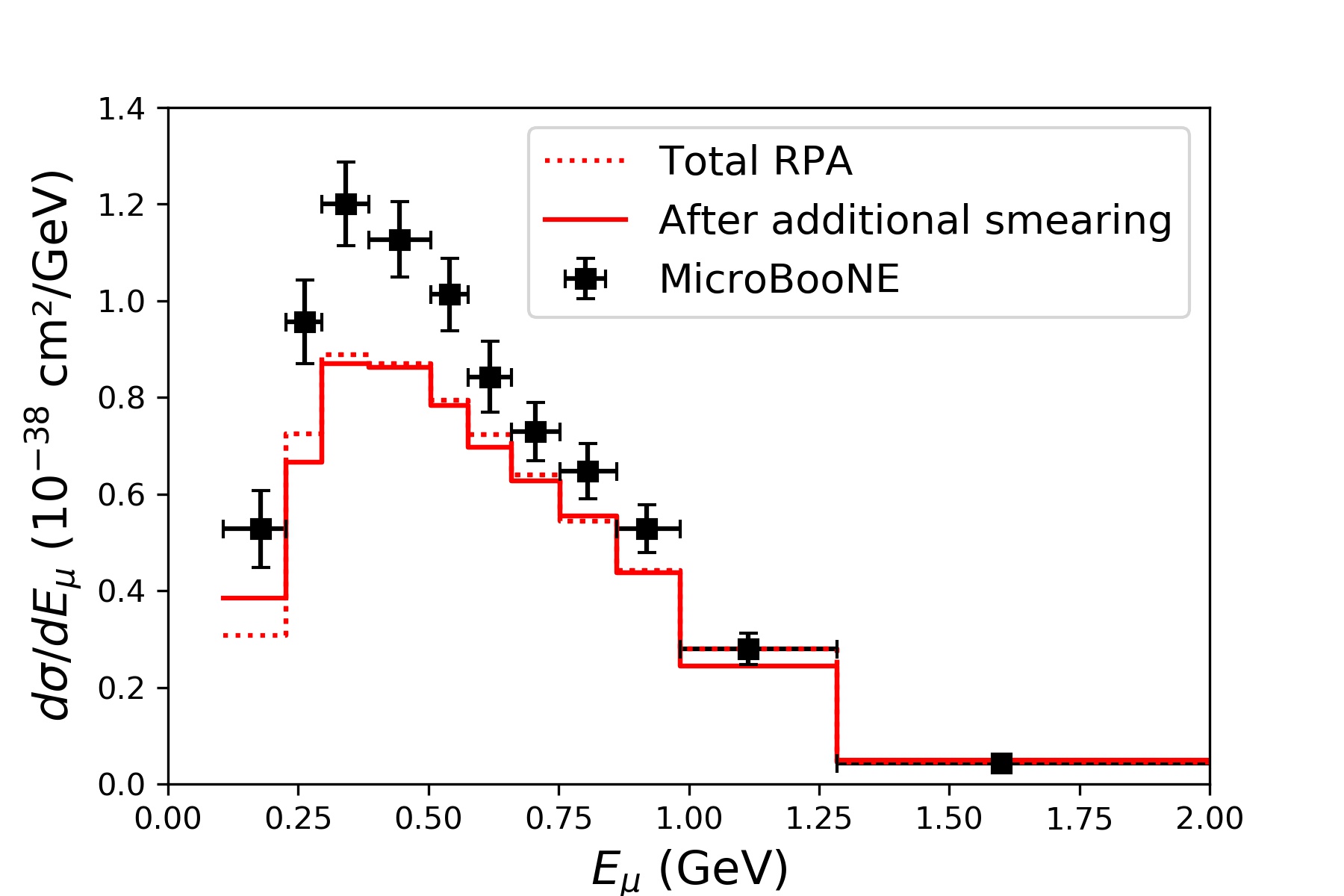}
\caption{Charged current inclusive MicroBooNE $\nu_\mu$ flux-integrated single differential cross
section on argon per nucleon as a function of the muon energy $E_\mu$.  
Left panel: the genuine quasielastic, the np-nh contributions, the coherent and incoherent 1$\pi$ production as well as the sum of these channels calculated in our RPA model are shown.
Right panel: our total RPA predictions averaged on different bins calculated before (dotted line) and after (continuous line) the additional MicroBooNE smearing explained in the text.
The experimental MicroBooNE data and the additional smearing matrix are taken from Ref.~\cite{MicroBooNE:2021cue}
}
\label{fig_dsigma_dEmu}
\end{center}
\end{figure}

\begin{figure}
\begin{center}
  \includegraphics[width=12cm]{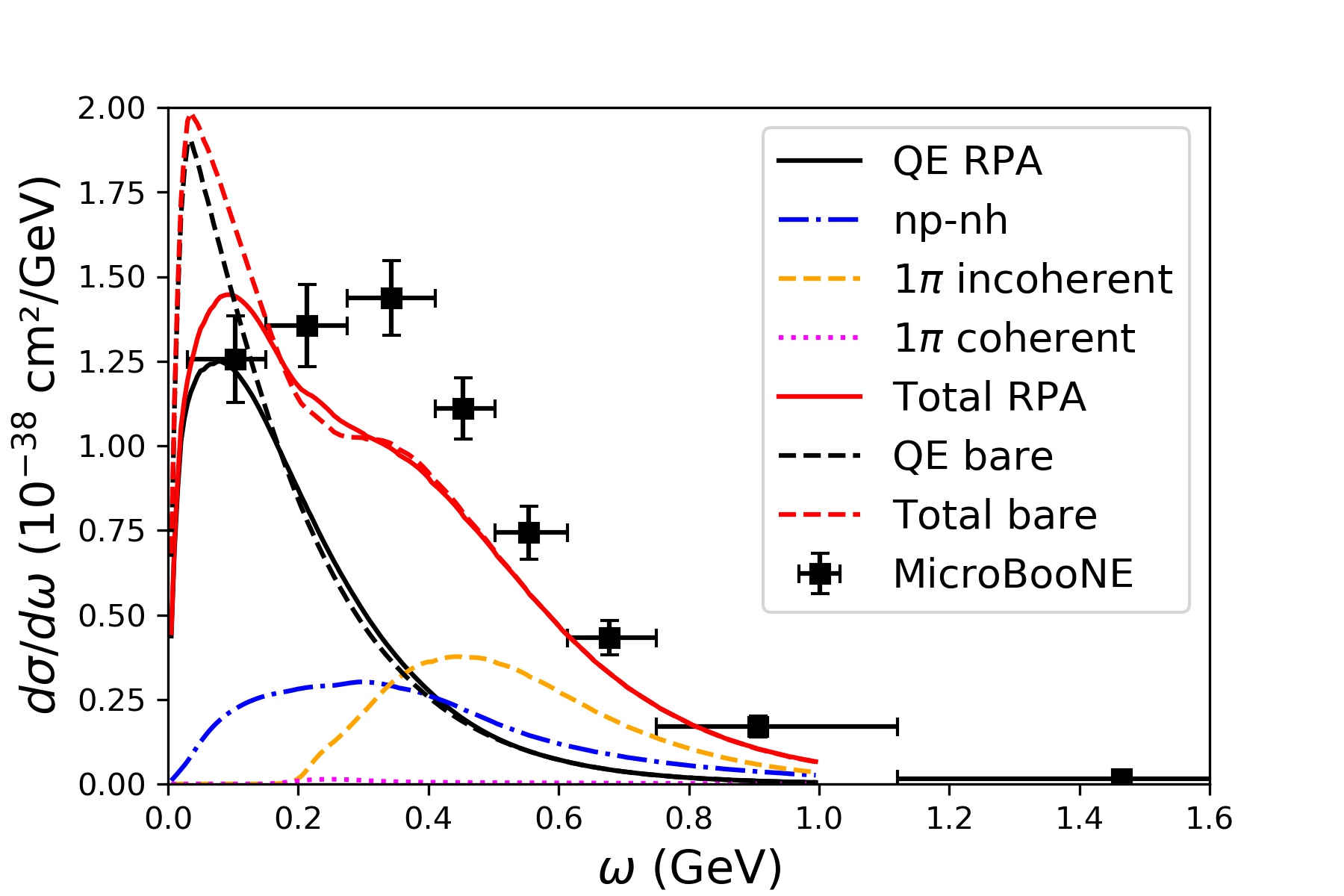}
\caption{Charged current inclusive MicroBooNE $\nu_\mu$ flux-integrated single differential cross
section on argon per nucleon as a function of the transferred energy $\omega$. The experimental MicroBooNE data are taken from Ref.~\cite{MicroBooNE:2021cue}.
The genuine quasielastic, the np-nh contributions, the coherent and incoherent 1$\pi$ production as well as the sum of these channels calculated in our RPA model are shown. 
The genuine quasielastic and the total cross section are shown also when the RPA is switched off.}
\label{fig_dsigma_domega}
\end{center}
\end{figure}

Concerning the total cross section as a function of the true neutrino energy $\sigma(E_\nu)$, we plot in Fig.\ref{fig_sigma_vs_Enu} the experimental inclusive result compared to our evaluation, with and without RPA effects, for the sum of all the channels included in our model (QE, np-nh, coherent and incoherent $1\pi$ production). 
Even if the published experimental results extend up to $E_\nu=2.6$ GeV, in Fig.\ref{fig_sigma_vs_Enu} we stop our comparison at $E_\nu=1.2$ GeV. The reason is that, as written in Ref.\cite{Martini:2009uj}, our numerical calculations of the nuclear responses, and consequently of the cross sections, extend only up to the transferred energy $\omega=1$ GeV. In the total cross section, such the one of Fig.\ref{fig_sigma_vs_Enu}, the $\omega >1$ GeV responses start to contribute for $E_\nu=1.2$ GeV. For the same reason, our theoretical curves of the flux-integrated $\frac{d\sigma}{d \omega}$ shown in Fig.\ref{fig_dsigma_domega} stop at $\omega=1$ GeV. On the contrary, this cut in the transferred energy is not a problem for the flux integrated $\frac{d^2\sigma}{d p_\mu~d\cos\theta}$ and $\frac{d\sigma}{d E_\mu}$ (shown in Figs. \ref{fig_comp_inclusive_micro} and \ref{fig_dsigma_dEmu} respectively)
for any $p_\mu$ or $E_\mu$ values, first since the $\omega>1$ GeV contributions are largely subdominant (almost negligible) in the kinematical conditions of the experimental results discussed here, second because an interpolation/extrapolation procedure to go from our initial input, the $\frac{d^2\sigma}{d\omega~d\cos\theta}$  at different fixed neutrino energies, to the flux-integrated $\frac{d^2\sigma}{dE_\mu~d\cos\theta}$ allows us to include these $\omega>1$ GeV  contributions. Returning to the total cross section as a function of the neutrino energy shown in Fig.\ref{fig_sigma_vs_Enu}, 
the agreement is good up to $E_\nu\simeq 0.7$ GeV. 
This is not the case of other models, 
such as GENIE v3 \cite{GENIE:2021npt}, MicroBooNE MC \cite{MicroBooNE:2021ccs}, NEUT \cite{Hayato:2021heg} and NuWro \cite{Golan:2012rfa} which underestimate the data,  with the exception of GiBUU \cite{Buss:2011mx}, as shown in Ref. \cite{MicroBooNE:2021cue}.
Beyond $E_\nu\simeq 0.7$ GeV our evaluation as well underestimates the data. A similar deviation between the data and our evaluation occurred for $^{12}$C cross section \cite{Martini:2014dqa} measured by SciBooNE \cite{Nakajima:2010fp}. This is due to inelastic channels missing in our description such as $2\pi$ production. 

A lack of strength is also visible in the  
flux integrated single differential cross section $\frac{d\sigma}{d E_\mu}$ plotted as a function of the muon energy $E_\mu$ in Fig.\ref{fig_dsigma_dEmu}. It appears in the same muon kinematical region as the one of the double differential cross section shown in Fig. \ref{fig_comp_inclusive_micro}. 
For $\frac{d\sigma}{d E_\mu}$ also the already mentioned models used in Ref.~\cite{MicroBooNE:2021cue} lead to an underestimation of the data. 
For a more quantitative comparison of our predictions with data and other models, we apply to our theoretical calculations the additional smearing needed for the comparison with the MicroBooNE data, as discussed and provided in the supplemental material of Ref. \cite{MicroBooNE:2021cue} via smearing matrices. This smearing is a result of the regularization in the data unfolding procedure. The results we obtain by applying this additional smearing are shown in the right panel of Fig.\ref{fig_dsigma_dEmu} 
which displays the smeared $\frac{d\sigma}{d E_\mu}$ cross section. The effect of the smearing is small. This will not be the case for the $\frac{d\sigma}{d\omega}$ cross section discussed later. 
Beyond the additional smearing matrices, the MicroBooNE collaboration also published the covariance matrices of
their cross section measurements. It allows to perform the $\chi^2$ test statistic via the covariance matrix formalism. We have calculated the $\chi^2/\textrm{ndf}$ of our model for the $\frac{d\sigma}{d E_\mu}$ cross section and we obtained $\chi^2/\textrm{ndf}=27.9/11$, a value larger than the one of GiBUU, $\chi^2/\textrm{ndf}$=16.7/11, the model which better predicts these data. However our $\chi^2/\textrm{ndf}$ is lower than the one of  GENIEv3 ($\chi^2/\textrm{ndf}$=32.4/11).

\begin{figure}
\begin{center}
  \includegraphics[width=8cm]{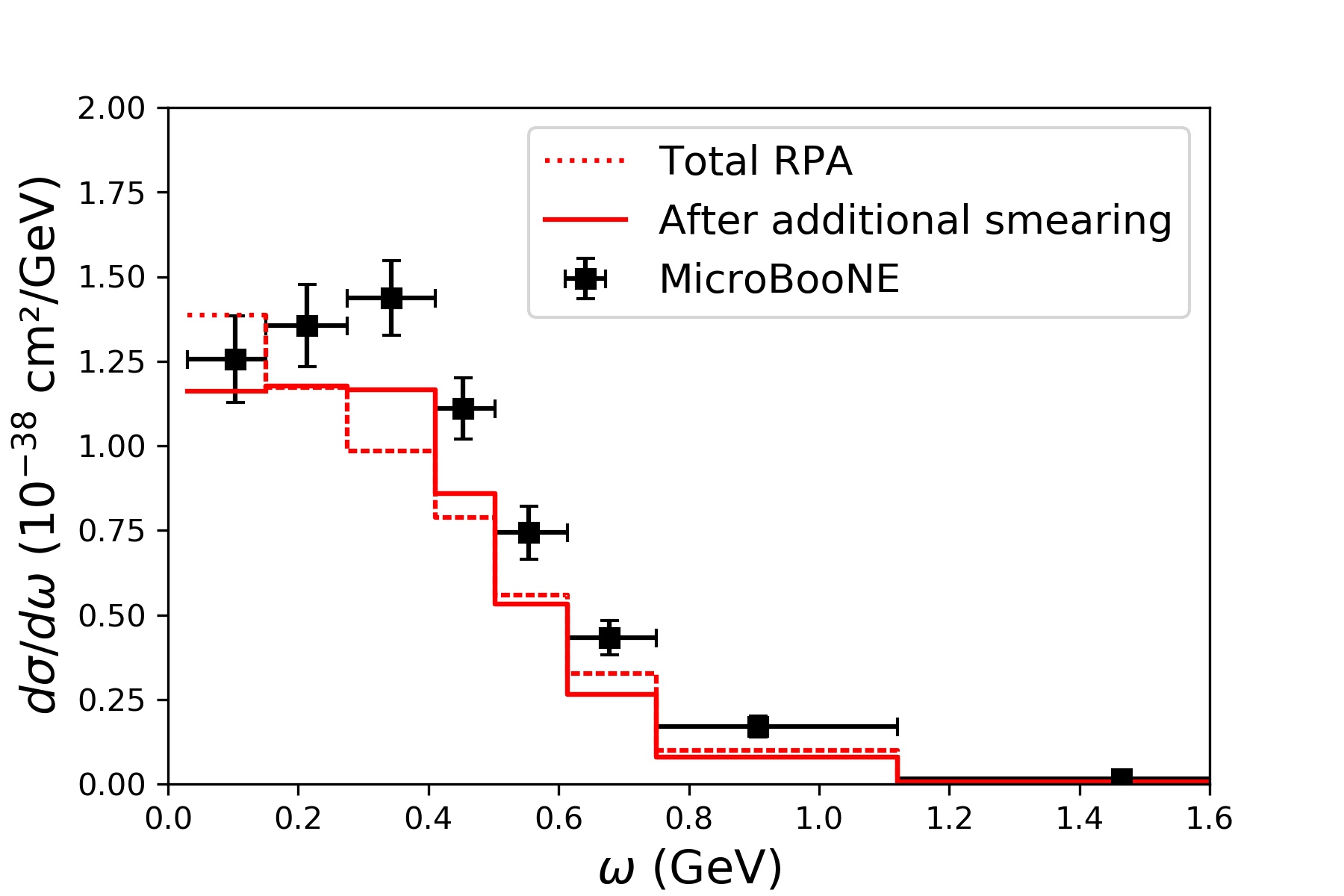}
\includegraphics[width=8cm]{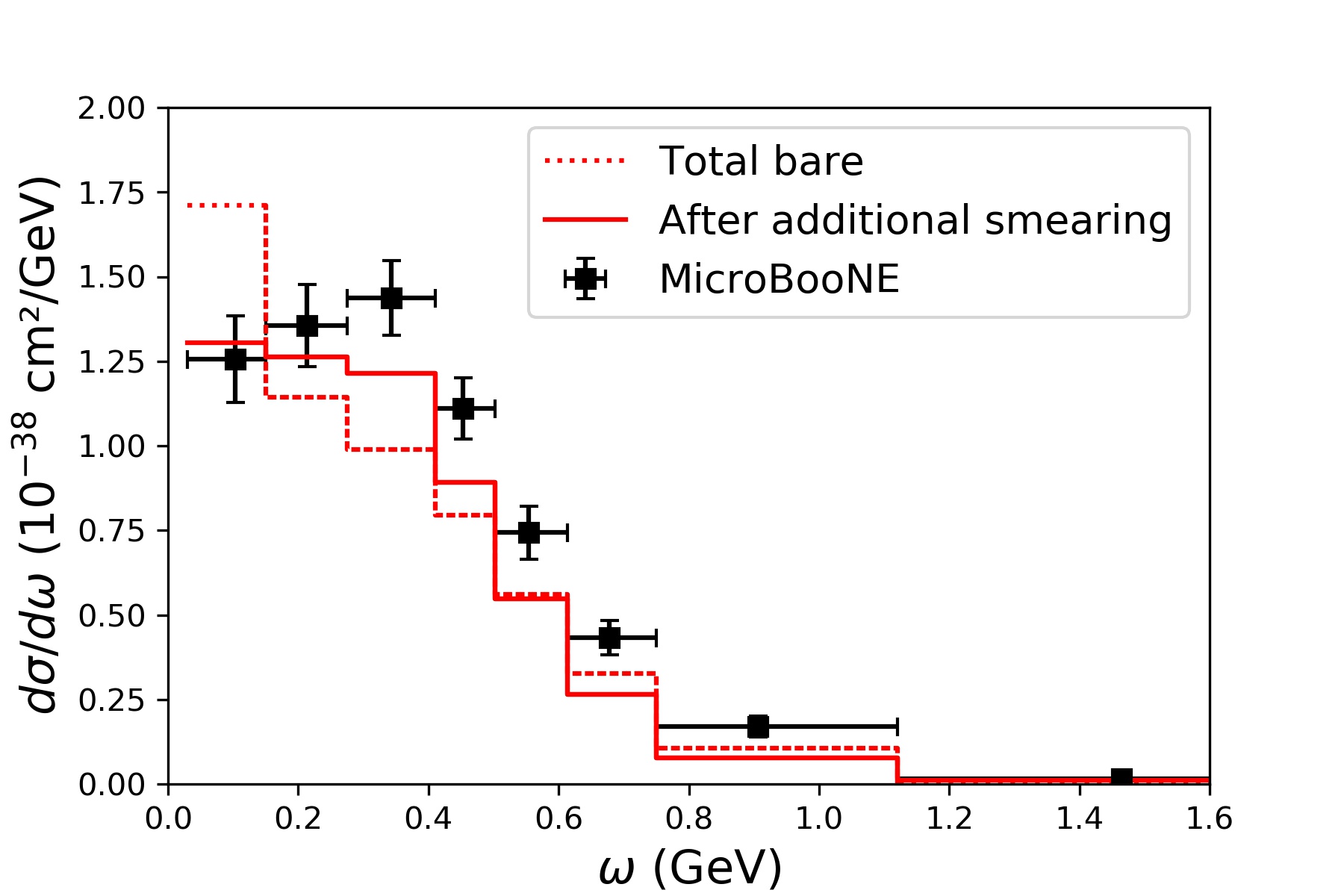}
\caption{Charged current inclusive MicroBooNE $\nu_\mu$ flux-integrated single differential cross
section on argon per nucleon as a function of the transferred energy $\omega$ calculated in our model  with (left panel) and without (right panel) RPA effects before (dotted lines) and after (continuous lines) the additional MicroBooNE smearing.
The experimental MicroBooNE data and the additional smearing matrix are taken from Ref.~\cite{MicroBooNE:2021cue}.
}
\label{fig_dsigma_domega_smeared}
\end{center}
\end{figure}

\begin{figure}
\begin{center}
  \includegraphics[width=12cm]{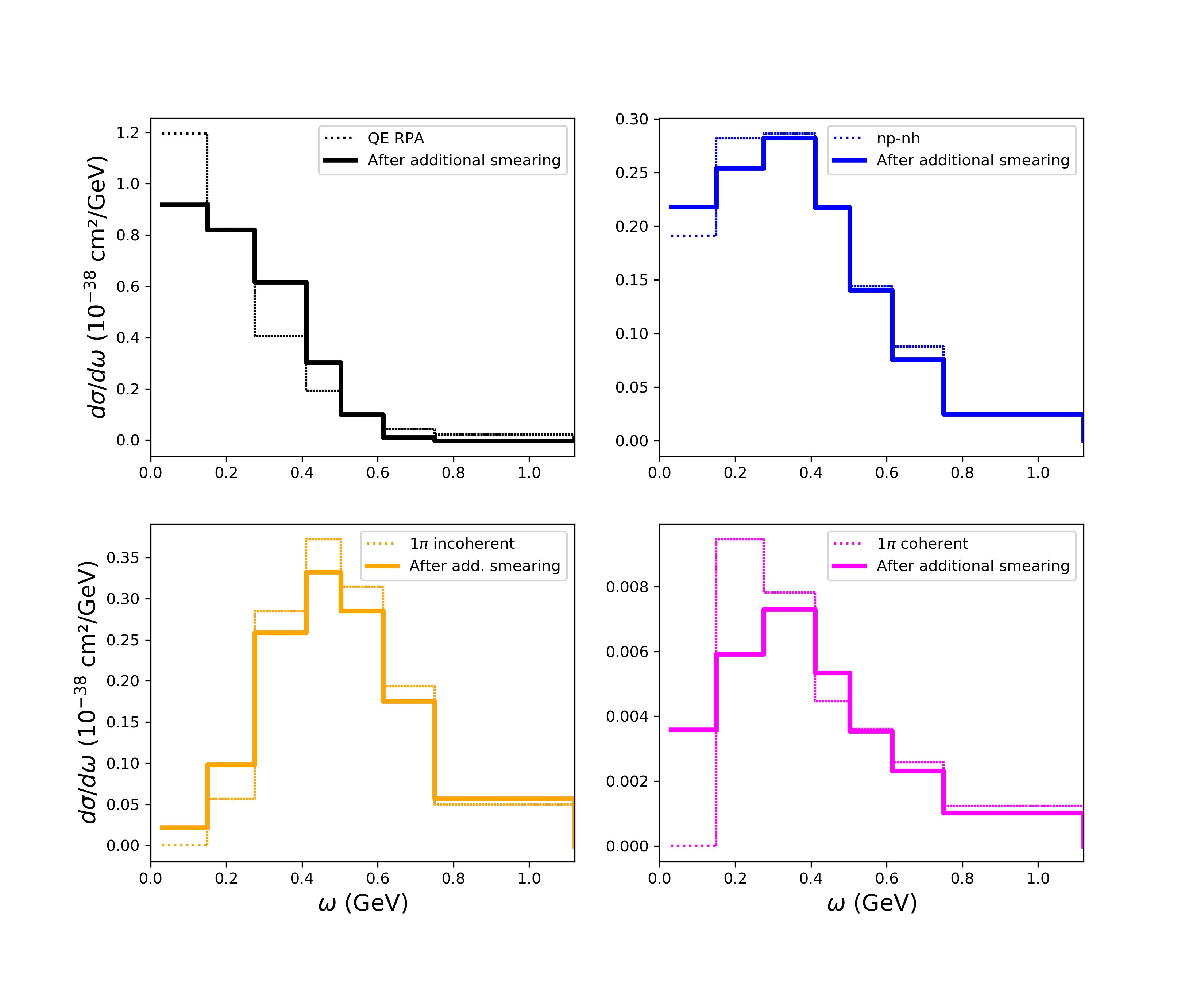}
\caption{Charged current 
quasielastic, multinucleon, coherent and incoherent 1$\pi$ production contributions to the MicroBooNE $\nu_\mu$ flux-integrated single differential cross
section on argon per nucleon as a function of the transferred energy $\omega$ 
calculated in our RPA model (dotted line) and by employing the additional MicroBooNE smearing (continuous line).}
\label{fig_dsigma_domega_smeared_channels}
\end{center}
\end{figure}

In figure \ref{fig_dsigma_dEmu}, which displays the differential cross section in terms of muon variables, the different reactions channels are entangled. 
The flux integrated differential cross section $\frac{d\sigma}{d\omega}$ as a function of the transferred energy $\omega$ allows instead a better separation of the different channel contributions, such as quasielastic, np-nh, 1$\pi$ production, which peak at different $\omega$ values. The comparison between the MicroBooNE data and our predictions for this cross section is shown in Fig.\ref{fig_dsigma_domega}.  
At low energy transfer the cross section is dominated by the quasielastic channel 
which is quenched by RPA effects in our theoretical calculations. 
 A lack of strength shows up for $0.2<\omega<0.6$ GeV but 
 the additional smearing should be applied to our curves before drawing any conclusions. This is done in Fig.\ref{fig_dsigma_domega_smeared} for the sum of all the channels we take into account,  with and without RPA effects. The first observation is that the 
impact of this smearing is larger for this $\frac{d\sigma}{d\omega}$ cross section than for the 
$\frac{d\sigma}{d E_\mu}$ one shown in 
Fig.\ref{fig_dsigma_dEmu}. 
The smearing produces a redistribution of the strength which is more important when the cross section is peaked, such as the quasielastic or the pion production, as illustrated also in Fig. \ref{fig_dsigma_domega_smeared_channels}. Furthermore this smearing reduces the difference between the results with and without RPA. 
 By using the uncertainty covariance matrix delivered by MicroBooNE for this $\frac{d\sigma}{d\omega}$ cross section as well, we calculate $\chi^2/\textrm{ndf}=17.2/8$ for our model including RPA effects. This result is comparable to the one of GiBUU, $\chi^2/\textrm{ndf}=17.0/8$, the model which better reproduces the data as compared to the other Monte Carlo predictions, characterized by a larger $\chi^2/\textrm{ndf}$, up to $\chi^2/\textrm{ndf}=33.8/8$ for MicroBooNE MC. 
A possible reason for our better agreement with data as compared to the predictions of GENIEv3, MicroBooNE MC,  NEUT and NuWro is that these models implement a multinucleon contribution deduced for the evaluation of Nieves \textit{et al.} \cite{Nieves:2011pp} which is smaller than our by about a factor 2 \cite{Morfin:2012kn,Katori:2016yel,Abe:2016tmq}. 
Ignoring RPA effects, our $\chi^2$ decreases to $\chi^2/\textrm{ndf}=13.4/8$. This is a consequence of the displacement of the QE strength from $\omega<0.2$ GeV to larger transferred energies due to the additional smearing, which partially compensates our underestimation of data for $\omega>0.2$ GeV. This underestimation which, even if less pronounced that in the other Monte Carlo predictions except GiBUU, remains even after the additional smearing, seems to start at $\omega \simeq 2 m_\pi$. It may signal the absence in our description of $2\pi$ production and other inelastic channels. This absence could also explain the underestimation of the inclusive MicroBooNE double differential cross section at low $p_\mu$, already pointed for Fig.\ref{fig_comp_inclusive_micro}. 
One wonders why this underestimation does not appear in the T2K inclusive data shown in Fig.\ref{fig_t2k_d2s_2018}. The reason is possibly due to the difference between the T2K and MicroBooNE neutrino energy profiles, the second one having a significant high energy contribution, absent for T2K, see Fig.\ref{fig_micro_t2k_fluxes}.

\section{Conclusions}
We have analyzed in this work the recent MicroBooNE data of neutrino cross sections on argon. We have considered the charged current inclusive 
measurements: the total cross section as a function of the neutrino energy and three flux integrated quantities, the double differential cross section as a function of the muon momentum and scattering angle, the single differential ones as a function of the muon energy and of the energy $\omega$ transferred to the nucleus. 
We have compared them to our theoretical approach. 
Overall we find an agreement with the data, 
in spite of a tendency of underestimation in some specific regions.  
The availability of covariant matrices for some experimental results allows quantitative comparisons between different models. 
Our model is particularly efficient in the case of the $\frac{d \sigma}{d \omega}$ data, a new type of measurement recently available. 
These data allow a better separation of the different reaction channels, even after the additional smearing needed for comparing models and data. Our analysis shows that the low $\omega$ region is dominated by the quasielastic. At larger $\omega$ a lack of agreement shows up: our predictions underestimate the data. The two pions production and other inelastic contributions which are not taken into account in our description are the natural candidates to explain this underestimation. These channels are more relevant for MicroBooNE than for T2K, due to the different energy profiles of these neutrino beams.

\section*{Acknowledgement}
We acknowledge the IPSA students B. Chaillié, V. Demoly and Q. Jacquet for the interest on this work in its early stage. We thank M. B. Barbaro, V. Belocchi, A. De Pace, T. Ericson, J. M. Franco-Patino, C. Giganti and W. Gu for interesting discussions. One of us (M. M.) acknowledges the support of the CERN Neutrino Platform and the 
hospitality of the CERN theory department, where part of
this work was done. 
\bibliography{bib_micro}
\end{document}